\documentclass[epjCONF]{svjour}
\usepackage{graphics}
\usepackage[varg]{txfonts} % Times fonts
\usepackage[latin1]{inputenc}
\usepackage[T1]{fontenc}
\usepackage{textcomp}
\usepackage{exscale}
\usepackage{amsfonts}
\usepackage{amssymb,amscd}
\usepackage[dvips]{epsfig}                   
\usepackage{array}
\usepackage{xcolor}
\usepackage{bbm}
\newcommand{\gS}[1]{#1\!\!\!\!\!\not~}	
\newcommand{\GS}[1]{#1\!\!\!\!\!\!\not~}

\newcommand{\pslash}{\gS{p}~}
\newcommand{\Pslash}{\GS{P}~}

\newcommand{\tr}{\textrm{tr}}
\newcommand{\ONE}{\mathbbm{1}}%{ {\bf 1} } 

\session-title{19th International IUPAP Conference on Few-Body Problems
in Physics}
\begin{document}
\title{Bethe--Salpeter studies of mesons beyond rainbow-ladder}
\author{Richard Williams\inst{1}\fnmsep\thanks{\email{richard.williams@physik.tu-darmstadt.de}}
}
\institute{Institute for Nuclear Physics, 
 Darmstadt University of Technology,
 Schlossgartenstra{\ss}e 9, 64289 Darmstadt, Germany 
 }
\abstract{
We investigate the masses of light mesons from
a coupled system of Dyson--Schwinger and Bethe--Salpeter equations. 
The dominant non-Abelian and sub-leading Abelian contributions
to the dressed quark-gluon vertex are explicitly taken into account.  We also include unquenching effects in the form of
hadronic resonance contributions via the back-reaction of pions.
We construct the corresponding Bethe--Salpeter 
kernel that satisfies the axial-vector Ward-Takahashi identity.
Our numerical treatment fully includes all momentum dependencies
with all equations solved completely in the complex plane.
This approach goes well beyond the rainbow-ladder approximation 
and permits us to investigate the relative impact of different
corrections beyond rainbow-ladder on the properties of mesons. We find
that sub-leading Abelian corrections are further suppressed dynamically,
and that our results supersede early qualitative predictions with
significantly simpler truncation schemes. 
} %end of abstract
\maketitle
\section{Introduction}\label{WilliamsR_intro}
In hard scattering experiments we can probe the structure of 
mesons and baryons, thus revealing their partonic building 
blocks; the quarks and gluons. This picture is natural at high-energies
where we have asymptotic freedom and mirrors the apparent degrees of 
freedom in the QCD Lagrangian. However, at low to intermediate energies
where the effects of confinement become more important, a description 
in terms of hadronic degrees of freedom is more appropriate. In the 
description of bound-states, in order to study their properties and 
interactions in terms of their constituent particles, we require 
non-perturbative tools that are capable of relating both the small 
and large scales together. This is necessary to provide, for example, a connection 
between the symmetries inherent in the QCD Lagrangian and their
manifestation in the hadronic spectrum.

Suitable tools include chiral perturbation theory, quark models and of 
course lattice QCD. Though heavy quark systems can be adequately described
with non-relativistic models, and lattice calculations are quickly 
approaching the physical point, we are also interested in how the light
meson spectrum arises from dynamical chiral symmetry breaking. This
suggests that we use an approach in which we can make a direct connection
to the limit of vanishing quark mass; moreover, we still wish to employ a description in 
terms of quarks and gluons rather than effective degrees of freedom. One 
such approach that is intrinsically non-perturbative, formulated in the 
continuum and capable of exploring both chiral and heavy quarks are the 
Dyson--Schwinger equations (DSEs) and Bethe--Salpeter equations (BSEs), 
which describe one-particle irreducible Green's functions and two-body 
bound states.

The DSE-BSE framework has been explored extensively in the literature
with regards to the light-meson sector over the last few decades, for
reviews see Refs.~\cite{Fischer:2006ub,Roberts:2007jh}. However,
the vast majority of studies have been undertaken under the umbrella of
the rainbow-ladder (RL) approximation, in combination with a
phenomenological approach to the quark-gluon interaction. The main
reason for the focus on this truncation is clear: firstly, it satisfies
the vector and flavour non-singlet axial-vector Ward-Takahashi identities
(axWTI) so necessary for the identification of the pion as a
(pseudo)Goldstone boson; and secondly because of its sheer simplicity. 
The last few years have seen these same studies extended to the description of
baryons via the quark-diquark approximation and 
more recently through the full covariant three-body Faddeev 
equations~\cite{Eichmann:2007nn,Eichmann:2008ef,Nicmorus:2008vb,Eichmann:2009qa}.

With the successes of RL in the meson sector~\cite{Maris:1997hd,Maris:1997tm,Maris:1999nt,Maris:2000sk,Bhagwat:2006pu}, and also for the baryons,
it is prudent to attempt to improve upon our approximation schemes. The 
most important object of interest here is the quark-gluon vertex since 
its specification ultimately determines the interaction between our 
fundamental degrees of freedom. An extension of the truncation scheme can
give rise to Yang-Mills corrections, unquenching effects and eventually
to coupled channel processes and decays; all of these are desirable properties that 
we wish to include in a more complete description of mesonic bound-states. 
Since a well-known prescription exists for constructing a DSE/BSE 
truncation based on diagrammatic expansions of the
vertex~\cite{Munczek:1994zz,Bender:1996bb}, this has been
the main focus over the last few years. Because of the sheer complexity 
of such studies which require higher loop integrals to be computed, most
studies~\cite{Bender:1996bb,Bender:2002as,Bhagwat:2004hn,Watson:2004kd,Matevosyan:2006bk}
employ the simplistic Munczek--Nemirovsky delta-function ansatz~\cite{Munczek:1983dx}
for at least one of the exchange gluons. This kills a loop integral making
studies tractable, and in some cases analytical, at the expense of barring 
the dominant vertex corrections from consideration. In some cases this
also restricts the study to just the pseudoscalars and vectors with
other mesons becoming unbound. We will later see that the use of such a
non-propagating ansatz gives rise to truncation artefacts that are hard
if not impossible to disentangle from the results, invalidating some of the
qualitative predictions such investigations give rise too.
Regardless of these concerns, past studies are pioneering in themselves
when one considers the resources available at the time with much being
learned as a result.

While the investigation of Bethe--Salpter equations is important, we must
also explore and improve our understanding and truncations of the
Dyson--Schwinger equations that provide their input. For example, the
Yang-Mills (YM) sector of QCD has also been much explored, resulting in
solutions of the ghost and gluon propagators that are in qualitative
agreement with lattice calculations, see Refs.~\cite{Fischer:2006ub,Fischer:2008uz}. These 
achievements encourage us
to steer away from phenomenological studies and attempt to construct higher
$n$-point Green's functions using all of the information
available~\cite{Schleifenbaum:2004id,Alkofer:2008dt,Kellermann:2008iw}. Recently,
Dyson--Schwinger studies of the quark-gluon vertex and quark
propagator have been made in which the dominant non-Abelian corrections and
sub-leading Abelian contributions were considered for the first time~\cite{Alkofer:2008tt}.
In lieu of these developments, and following our programme of
investigating Bethe--Salpeter equations beyond
rainbow-ladder~\cite{Fischer:2008sp,Fischer:2008wy}, a significant extension of the
Bethe--Salpeter equation beyond the rainbow-ladder truncation has been
proposed~\cite{Fischer:2009jm}. There, the dominant non-Abelian
corrections to the quark-gluon vertex are considered in a symmetry 
preservation truncation of the Bethe--Salpeter equation, with the
corresponding light-meson spectrum calculated. 

Because of the technical challenges in such a state-of-the-art calculation 
the study was exploratory, relying upon phenomenological input for the gluon
propagator. Though this is undesirable in light of what we know about
the basic Green's functions of QCD, this does not trivialize the calculation 
itself. The integrals in Euclidean space remain two-loop in nature, and require 
our input quantities to be evaluated for complex momenta. The aim of the
present paper is not to take the next step and introduce a quantitatively reliable truncation
scheme. Rather, our intention is provide further  corrections to the quark-gluon
vertex in addition to those considered thus far~\cite{Fischer:2009jm}.
Hence, we will include both the sub-leading Abelian corrections and 
the pion back-reaction that represents hadronic unquenching
effects. As a result, we obtain a model that is orders of magnitude 
more complicated than the monumental step already made. Thus, for the time-being, 
we satisfy ourselves with a qualitative 
study of corrections beyond rainbow-ladder and their relative impact on
mesonic bound states.  The extension to a consistent
and qualified truncation scheme will be reported in detail elsewhere.

This paper is organised as follows: in section \ref{WilliamsR_sec:dse} we introduce the
Dyson--Schwinger equation for the quark-propagator and the quark-gluon vertex.
The truncation scheme is introduced, being sub-divided into a non-resonant 
Yang-Mills part and a resonant hadronic contribution. In section \ref{WilliamsR_sec:bse}
we introduce the Bethe--Salpeter equation for a quark-antiquark
bound-state, detailing the 
relativistic decomposition of the bound-state amplitude, general
solution methods, their normalisation and the interaction kernel. In 
section \ref{WilliamsR_sec:results} we give the results of our calculation and compare 
to existing literature. Finally, we conclude and provide an outlook in 
section \ref{WilliamsR_sec:outlook}.

\section{Dyson--Schwinger Equations\label{WilliamsR_sec:dse} }

\subsection{Quark propagator}
\begin{figure*}[t]
\centerline{\includegraphics[width=0.7\textwidth]{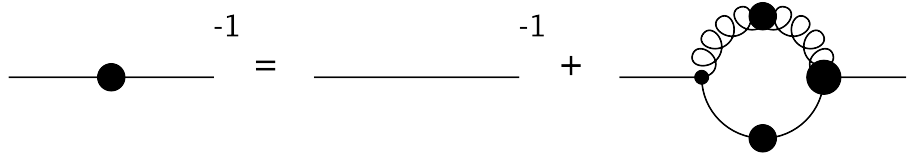}}
\caption{The Dyson--Schwinger equation for the fully dressed quark
propagator. Wiggly lines represent gluons and straight lines quarks. Large
filled circles indicate the quantity is fully-dressed, otherwise it is bare.}\label{WilliamsR_fig:quarkdse}
\end{figure*}
The quark propagator and the gap equation it satisfies is one of the
most important quantities in the covariant description of mesons. It encodes such
non-perturbative properties as dynamical mass generation and the
realisation of a non-zero condensate through the dynamical breaking of
chiral symmetry.  Moreover, as we will see in
section~\ref{WilliamsR_sec:cutting} a symmetry preserving truncation of the
Bethe--Salpeter kernel can be constructed directly from the self-energy
part of the quark Dyson--Schwinger equation.

In Euclidean momentum space, the renormalised dressed gluon and 
quark propagators in the Landau gauge are given by
\begin{eqnarray}
  D_{\mu\nu}(p) &=&  \left( \delta_{\mu\nu}-\frac{p_\mu p_\nu}{p^2}  
  \right)\frac{Z(p^2;\mu^2)}{p^2}           \,,\label{WilliamsR_eqn:gluon}\\[2mm]
  S(p) &=&  \frac{Z_f(p^2;\mu^2)}{i\pslash  + M(p^2)} =
  \frac{1}{i\pslash A(p^2;\mu^2)+B(p^2;\mu^2)}\,,  \label{WilliamsR_eqn:quark}
\end{eqnarray}
where $Z(p^2;\mu^2)$ is the gluon dressing function, $Z_f(p^2;\mu^2)$ is the
quark wave-function and $M(p^2)$ is the renormalisation point independent 
quark mass function. The dependence of such functions on the renormalisation 
point $\mu^2$ will be implicitly assumed from here on. The quark dressing 
functions $A(p^2)$ and $B(p^2)$ can be recombined into the quark mass and 
wave-function by $M(p^2) = B(p^2)/A(p^2)$ and $Z_f(p^2) = 1/A(p^2)$.

These propagators may be obtained by solving their respective
Dyson--Schwinger equations. The DSE for the quark propagator, shown
diagrammatically in Fig.~\ref{WilliamsR_fig:quarkdse}, is written
\begin{eqnarray}
	S^{-1}(p) &=& Z_{2}S^{-1}_{0}(p)
	+\Sigma(p)\;,\nonumber\\[-2mm]\label{WilliamsR_eqn:quarkdse}\\[-2mm]\nonumber
	\Sigma(p) &=& g^{2}C_{F}Z_{1F}\int\frac{d^4q}{\left( 2\pi \right)^4}
	\Gamma_{\nu}(q,p)  D_{\mu\nu}(k) 
	\gamma_{\mu}S(q)\;,
\end{eqnarray}
where $\Sigma(p)$ is the quark self-energy, $k-p-q$ and the Casimir $C_F=4/3$ stems from the colour trace. We
introduced the reduced quark-gluon vertex $\Gamma_\nu(q,p)$
defined by $\Gamma_\nu^a(q,p)=ig \frac{\lambda^a}{2}\Gamma_\nu(q,p)$. The bare
inverse quark propagator is $S^{-1}_{0}(p) = i \pslash + m$. The
renormalisation
factors are $Z_{1F}=Z_2/\widetilde{Z}_3$ for the quark-gluon vertex, $Z_2$
for the quark propagator and $\widetilde{Z}_3$ for the ghost dressing function.

The scalar dressing functions of the quark DSE are solved for by
appropriate projections of Eq.~(\ref{WilliamsR_eqn:quarkdse}). This is a
coupled non-linear integral equation that is solvable provided we know
the gluon dressing function and the structure of the quark-gluon vertex.
In the rainbow approximation both are specified by Ans\"atze. Here, the
quark-gluon vertex will be provided by solving its respective DSE in
a truncation scheme to be introduced in the next section.
Following the lead of past studies and our own investigations we employ the
momentum dependent ansatz~\cite{Alkofer:2002bp}
\begin{equation}
  Z(q^2) = \frac{4\pi}{g^2}\frac{\pi D }{\omega^2}\;q^4\;
  e^{-q^2/\omega^2}\;,\label{WilliamsR_eqn:watson}
\end{equation}
which is a Gaussian distribution whose scale and strength of the effective
gluon interaction are provided by $D=16$ and $\omega=0.5$, respectively.  Since it cuts out
the dynamics at large momenta all of the equations are finite and do not
require renormalisation. Such factors may then be set to one, though for
reference we still show them explicitly. The use of such an interaction 
with this simple (but non-trivial) form simplifies some aspects of the 
calculation which will remain somewhat involved.

\begin{figure*}[t!]
\centering
\begin{eqnarray}
	\mathrm{(a)}\begin{array}{c}
\includegraphics[width=0.7\textwidth]{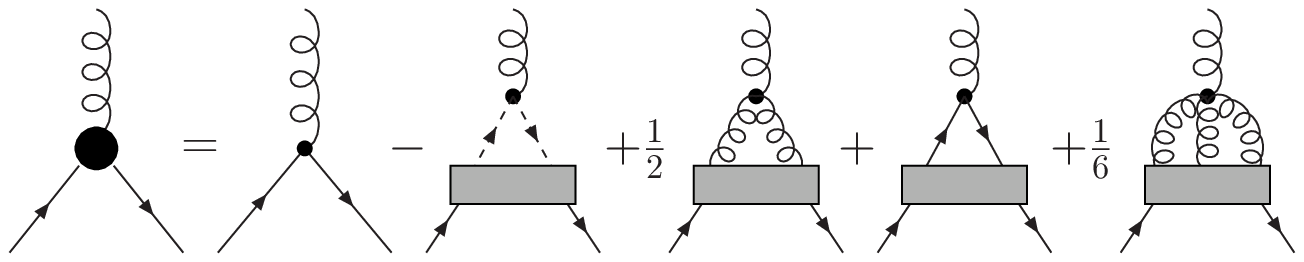}
\end{array}\label{WilliamsR_dse1}\nonumber\\
\mathrm{(b)}\begin{array}{c}
\includegraphics[width=0.7\textwidth]{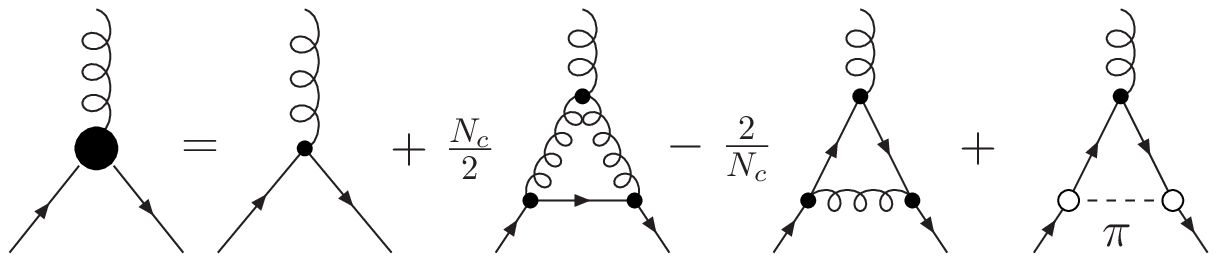}
\end{array}\label{WilliamsR_dse2}\nonumber
\end{eqnarray}
\caption{
In (a) we show the full Dyson--Schwinger equation for the quark-gluon vertex, and
(b) the truncation followed in this work.  All internal
propagators are dressed, with wiggly lines indicating gluons, straight
lines quarks and dashed lines mesons. White-filled circles indicate
bound-state amplitudes whilst black-filled ones represent vertex dressings.
Note that the last diagram of (b) is also proportional to $1/N_c$ and is
further approximated in what follows.
\label{WilliamsR_fig:qgdse}}
\end{figure*}

\subsection{Quark-Gluon vertex}
Though the quark propagator is one of the most important inputs to our
Bethe--Salpeter equation, the most important quantity with regards to the
dynamics of the theory is the quark-gluon vertex. This describes the
non-perturbative interplay between our dressed quarks and gluons. We are
left with two choices as to how to specify the details of this
interaction: either we
provide some physically motivated Ansatz for the
quark-gluon vertex; or we attempt to solve its corresponding Dyson--Schwinger 
equation, shown in full in Fig.~\ref{WilliamsR_fig:qgdse}(a). Since \emph{a
priori} we don't know how the quark-gluon behaves non-perturbatively,
the provision of an ansatz is challenging. The most well-studied QFT in
the literature with regards to vertex functions and the construction of
ans\"atze is QED; they usually employ restrictions from the Ward-Takahashi
identities and multiplicative
renormalisability~\cite{Ball:1980ay,Curtis:1990zs,Kizilersu:2009kg}.  
However, since QCD is by no means an Abelian theory it is not clear how
much information can be directly transferred between gauge
theories~\cite{Skullerud:2003qu,LlanesEstrada:2004jz,Kizilersu:2006et}.

One thing we do know for sure is how to decompose our vertex into an
appropriate basis.  In general, the quark-gluon vertex can be written as twelve
independent tensors built-up of two independent four-momenta $p_1^\mu$,
$p_2^\mu$ and the Dirac matrices $\ONE$, $\gamma^\mu$. An oft used basis has been provided for us by Ball and Chiu~\cite{Ball:1980ay}. The twelve components are split into
longitudinal and transverse parts
\begin{equation}
  \Gamma^\mu = \sum_{i=1,4} \lambda_i L_i^\mu 
             + \sum_{i=1,8} \tau_i    T_i^\mu\;,
  \label{WilliamsR_eqn:ballchiu}
\end{equation}
where the longitudinal and transverse basis components $L^\mu$, $T^\mu$
in Euclidean space are given in Table~\ref{WilliamsR_tab:ballchiu}.
\begin{table}[h]
	\begin{center}
\begin{eqnarray}
L_1^\mu&=& \gamma^\mu \nonumber\\ \nonumber
L_2^\mu&=&-(\gS{p}_1 + \gS{p}_2)(p_1+p_2)^\mu \\  \nonumber
L_3^\mu&=&-i(p_1+p_2)^\mu \\  \nonumber
L_4^\mu&=&-i \sigma^{\mu \nu}(p_1+p_2)_\nu \\  \nonumber
T_1^\mu&=& i(p_1^\mu p_2\cdot p_3-p_2^\mu p_1 \cdot p_3) \\  \nonumber
T_2^\mu&=&(p_1^\mu p_2\cdot p_3-p_2^\mu p_1 \cdot p_3)(\gS{p}_1+\gS{p}_2)) \\  \nonumber
T_3^\mu&=& \gS{p}_3 p_3^\mu -p_3^2 \gamma^\mu \\  \nonumber
T_4^\mu&=& -i(p_3^2 \sigma^{\mu \nu}(p_1+p_2)_\nu +2p_3^\mu
\sigma_{\lambda \nu} p_1^\lambda p_2^\nu ) \\  \nonumber
T_5^\mu&=& i \sigma^{\mu \nu} (p_3)_\nu \\  \nonumber
T_6^\mu&=&  (p_1^2-p_2^2)\gamma^\mu + (p_1+p_2)^\mu \gS{p}_3 \\ \nonumber
T_7^\mu&=& \frac{i}{2}(p_1^2-p_2^2) [(\gS{p}_1+\gS{p}_2)\gamma^\mu -(p_1+p_2)^\mu] \\ \nonumber
&&-i (p_1+p_2)^\mu \sigma_{\lambda \nu} p_2^\lambda   
p_1^\nu\\  \nonumber
T_8^\mu&=& -\gamma^\mu \sigma_{\lambda \nu} p_2^\lambda p_1^\nu
-\gS{p}_2 p_1^\mu +\gS{p}_1 p_2^\mu\;.
\end{eqnarray}
\caption{Longitudinal and transverse components of the quark-gluon
vertex in the Ball--Chiu basis. Here, the incoming gluon momentum
is $p_3=p_2-p_1$ with $p_1$, $p_2$ the incoming and outgoing quark
momentum. We use $\sigma_{\mu \nu}=\frac{1}{2}(\gamma_\mu \gamma_\nu - \gamma_\nu
\gamma_\mu)$.\label{WilliamsR_tab:ballchiu}}
\end{center}
\end{table}
The scalar coefficients that parametrise this vertex,
$\lambda_i$ and $\tau_i$, are in general functions of $p_1^2$, $p_2^2$,
and $p_3^2$. We note that this basis is only favoured for comparative
purposes. Since we work in Landau gauge where the gluon
propagator is explicitly transverse (and always accompanies the
quark-gluon vertex), a reduced basis of just eight components is
necessary. However, the choice of basis is irrelevant to the final
solution though a different basis can provide significant simplifications 
to the calculation.

Rather than attempting to provide an ansatz for the quark-gluon vertex,
we choose a more systematic approach and attempt to solve its 
Dyson--Schwinger equation. Since this is an intricate and highly non-trivial
coupled integral equation, containing basically unknown four- and five-point
functions, we must impose a truncation scheme. Thus, we approximate the
DSE following the detailed investigation of Refs.~\cite{Fischer:2007ze,Alkofer:2008tt}. 
This results in the approximate DSE portrayed in Fig.~\ref{WilliamsR_fig:qgdse}(b)
in which two-loop diagrams have also been neglected.
Here, the first \emph{non-Abelian} diagram subsumes the first two diagrams in
the full DSE to first order in a dressed skeleton expansion of the
four-point functions. The quark-antiquark kernel in the third loop
diagram of Fig.~\ref{WilliamsR_fig:qgdse}(a) is expanded in terms of resonance
contributions to the kernel and one-particle irreducible Green's
functions~\cite{Fischer:2007ze}. This gives rise to the second 
\emph{Abelian} diagram of Fig.~\ref{WilliamsR_fig:qgdse}(b) and one that involves
pion exchange.

In Fig.~\ref{WilliamsR_fig:qgdse}(b) we have shown explicitly the colour factors 
of the \emph{non-Abelian} and \emph{Abelian} diagrams. While the non-Abelian
diagram is associated with colour factors $N_c/2$, the Abelian diagram
is $N_c^2$ suppressed with a colour factor of $-2/N_c$. Similarly, the
resonant contribution is also proportional to $1/N_c$ due to  
the implicit $1/\!\sqrt{N_c}$ dependence of the two pion amplitudes. This
gives us an indication as to the relative strength of the contributions
from each diagram. It is clear that in the approximation scheme considered 
here all corrections are sub-dominant compared to the diagram containing 
the three-gluon vertex.

In Ref.~\cite{Fischer:2009jm} we discarded the sub-leading corrections, 
keeping only the non-Abelian diagram. Here we provide an extension by 
including the sub-dominant diagrams and investigate their contributions
in turn. For distinction we will classify our diagrams as follows: the non-resonant
non-Abelian and Abelian diagrams will be referred to collectively as the 
Yang-Mills part; the remaining resonant diagram corresponding to
unquenching effects will be addressed as the pion back-coupling.
In other words, we separate the hadronic resonance contributions from 
the other corrections to the quark-gluon vertex. We introduce each of these in
turn in the following.

\subsubsection{Yang-Mills contribution}
The corrections beyond rainbow-ladder that correspond to non-resonant
corrections will now be considered. In order to keep our calculation tractable 
we employ the well-established strategy of absorbing all internal vertex dressings
of the diagram into effective dressing functions for the two
internal gluon propagators. The resulting Dyson--Schwinger equation for the 
quark-gluon vertex $\Gamma^{\mu}(p_1,p_2)$ with quark momenta $p_1$ 
and $p_2$ and gluon momentum $p_3$ reads 
\begin{eqnarray} 
  \Gamma^{\mu}(p_1,p_2) &=& Z_{1F} \gamma^\mu+ 
\left( \frac{-i N_c}{2} g^2 Z_{1F}^2 Z_3\right)\nonumber \\ 
&&\hspace{-18pt}\times \int\nolimits_q 
\bigg\{  \gamma^\nu S(q)
\gamma^\rho 
\Gamma^{\rm 3g}_{\sigma\theta\mu}(k_1,k_2)
D_{\nu\sigma}(k_1) D_{\rho\theta}(k_2)
\bigg\} \nonumber\\
&&\hspace{-55pt}+\left( \frac{1}{2 N_c}g^2 \right)\int\nolimits_q\bigg\{ \gamma^\nu
S(k_1)\gamma^\mu S(k_2)\gamma^\rho D_{\nu\rho}(q)
\bigg\}\,.\label{WilliamsR_eqn:qg}
\end{eqnarray}
with $\int_q \equiv \int \frac{d^4q}{(2\pi)^4}$, the renormalisation 
factors $Z_{1F}$,$Z_1$ and $\Gamma^{\rm 3g}$ the bare three-gluon vertex. 

For the gluon, we employ the momentum dependent ansatz Eq.~(\ref{WilliamsR_eqn:watson}).
Naturally, such an ansatz provides only a first step towards a full calculation 
of the Abelian and non-Abelian diagrams, including input from the DSEs for the 
three-gluon vertex and the gluon propagator. Nevertheless we believe that the 
ansatz of Eq.~(\ref{WilliamsR_eqn:watson}) is sufficient to provide for reliable 
qualitative results as concerns the relative effects of vertex
corrections onto meson properties. In particular it is not sensitive to the question
of scaling vs. decoupling~\cite{Fischer:2008uz} in the deep infrared, $p<50$ MeV: 
both scaling and decoupling lead to a combination of three-gluon vertex and gluon propagator dressings that is vanishing in the infrared in qualitative agreement 
with the ansatz Eq.~(\ref{WilliamsR_eqn:watson}). In addition, quantitative effects in 
the interaction below $p<50$~MeV are not expected to affect observables in the flavour
non-singlet sector since the dynamical mass of the quark, 
$M \approx 350$~MeV, suppresses all physics on scales $p \ll M$ (see 
however \cite{Alkofer:2008et}). Finally, the ansatz Eq.~(\ref{WilliamsR_eqn:watson}) is 
not sensitive to details of the Slavnov--Taylor identity (STI) for the three-gluon
vertex: exactly those longitudinal parts of the vertex that are constrained from 
the STI are projected out in any Landau gauge calculation by the attached 
transverse gluon propagators.

\subsubsection{Unquenching Effects}

\begin{figure}[h!]
  \begin{center}
\begin{eqnarray*}
	\begin{array}{c}\vspace{-0.5cm}
	\includegraphics[scale=0.42]{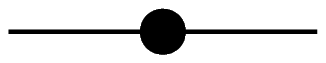}\\
    \end{array}^{-1}
	&=&
	\begin{array}{c}\vspace{-0.3cm}
	\includegraphics[scale=0.42]{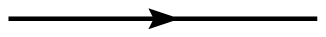}\\
	\end{array}^{-1}
	+
	\begin{array}{c}
	\includegraphics[scale=0.42]{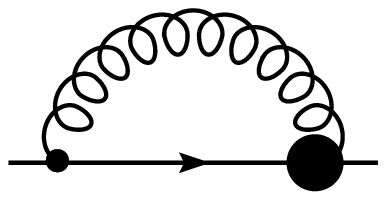}\\
	\end{array}
	+
	\begin{array}{c}
	\includegraphics[scale=0.42]{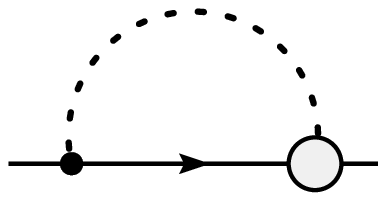}\\
	\end{array}
\end{eqnarray*}
\caption{Quark DSE including pion unquenching effects\label{WilliamsR_fig:unquench}}
\end{center}
\end{figure}

In the previous section we introduced the leading and sub-leading
non-resonant corrections to the quark-gluon vertex. Here we consider
the dominant hadronic resonant contributions corresponding to pion
exchange. The contribution is further approximated according to
~\cite{Fischer:2007ze,Fischer:2008sp} such that a symmetry preserving
Bethe--Salpeter kernel is constructable. This results in the
one-loop pion exchange contribution to the quark propagator, as shown in
Fig.~\ref{WilliamsR_fig:unquench}.  The DSE for the quark, Eq.~(\ref{WilliamsR_eqn:quarkdse}) is modified accordingly
\begin{eqnarray}
	S^{-1}(p) &=& Z_{2}S^{-1}_{0}(p)  +
	g^{2}C_{F}Z_{1F}\int\nolimits\frac{d^4q}{\left( 2\pi \right)^4}
	\gamma_{\mu}S(q)\nonumber \\
	&&\hspace{3cm}\times\,\Gamma_{\nu}(q,p)  D_{\mu\nu}(k)   \nonumber\\
	&& -3\int\nolimits\frac{d^4q}{\left( 2\pi \right)^4}\bigg[ Z_2 \gamma_5
	S(q)\Gamma_\pi\left( \frac{p+q}{2};-k \right)\nonumber\\
	&& \hspace{1.62cm}+Z_2\gamma_5
	S(q)\Gamma_\pi\left( \frac{p+q}{2};k  \right)
	\bigg]\frac{D_\pi}{2}\,,\label{WilliamsR_eqn:unquenchdse}
\end{eqnarray}
where $D_\pi(p) = 1/\left(p^2+M_\pi^2\right)$ is the pion propagator.
The factor of three in the pion contribution comes from the flavour
trace and represents contributions from $\pi^{0,\pm}$ which we treat
equally in the isospin limit.

To further reduce the complexity of such a contribution, we employ the simple
off-shell prescription for the pion used in~\cite{Fischer:2008wy} where we take 
only the leading component  of the pion amplitude in the exchange
diagram. Moreover, we replace it by its exact chiral limit value
\begin{equation}
F_1(p;P) = B_\chi(p^2)/f_\pi\;,
\end{equation}
where $B_\chi(p^2)$ is the scalar quark dressing function in the chiral
limit to ensure the correct asymptotic behaviour, and $f_\pi$ is
the leptonic decay constant. This ensures that the Bethe--Salpeter kernel has a non-trivial
momentum dependence are produces results that are qualitatively in
accord with phenomenology.

\subsection{Solution}
The system of Dyson--Schwinger equations for the quark-gluon
vertex and quark propagator are solved by applying suitable
projectors and taking the Dirac trace. This gives rise to fourteen
coupled integral equations for the (scalar) basis coefficients shown in
Eqs.~(\ref{WilliamsR_eqn:quark}) and (\ref{WilliamsR_eqn:ballchiu}) which we solve for on a discretised momentum
and angular grid via gaussian quadrature. This can be accomplished very
efficiently and without requiring large computing resources. For the
Bethe--Salpeter equation it becomes necessary to evaluate the quark
propagator, and hence the quark-gluon vertices at complex values of the
Euclidean momenta. By choosing the momentum routing
in an appropriate fashion, we can avoid the unconstrained  
analytic continuation of the gluon propagator and three-gluon 
vertex. To obtain solutions, we employ the `shell-method' described in the 
appendix of Ref.~\cite{Fischer:2007ze}.

In Fig.~\ref{WilliamsR_fig:quarkfn} we show the quark wave-function and
mass-function for different truncations of the quark-gluon vertex: RL is
rainbow-ladder; NA is the non-Abelian diagram; AB is the Abelian
diagram; and PI represents the pion back-reaction.

\begin{figure}[h!]
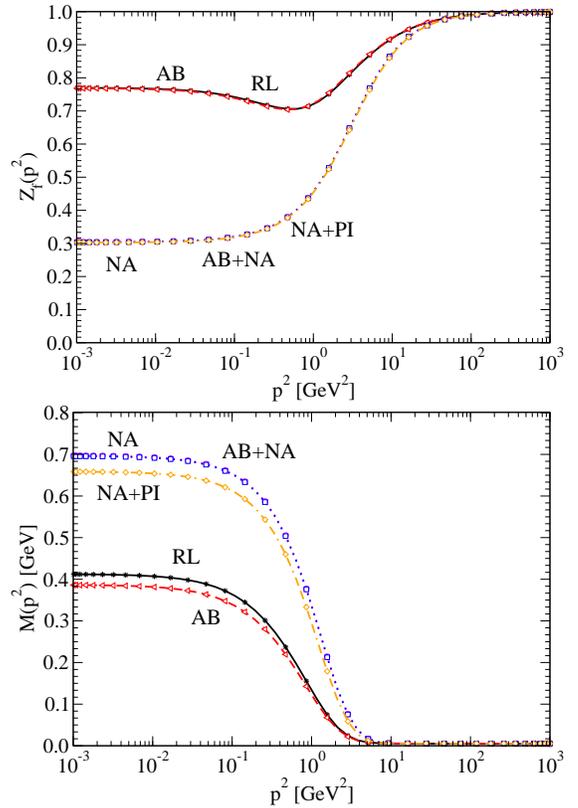

	\begin{center}
   \includegraphics[width=0.9\columnwidth]{WilliamsR-fig8.eps}
   \includegraphics[width=0.9\columnwidth]{WilliamsR-fig9.eps}
   \end{center}
\caption{\label{WilliamsR_fig:quarkfn}
  The mass function $M(p^2)=B(p^2)/A(p^2)$ of the quark and the wave 
  function $Z_f(p^2)$ for rainbow-ladder (RL), Abelian (AB),
  non-Abelian (NA), Abelian + non-Abelian (AB+NA) and non-Abelian + pion
  back-reaction (NA+PI).}
\end{figure}

As can be seen, inclusion of the non-Abelian diagram leads to a much
larger change in the mass function than the Abelian diagram alone.
Taking both diagrams together it is obvious that the effect of the Abelian
diagram is largely suppressed and is hidden underneath the NA curve
(the difference in the mass-function is approximately $1$~MeV at zero
momenta). 
The only significant contributions arises when the pion back-reaction
is also included, giving rise to the customary reduction of the quark mass-function.
However, although the Abelian contribution appears suppressed here its
impact on the Bethe--Salpeter equation must still be investigated since
the dynamics associated with crossed-ladder kernels are often expected
to be important.
\section{Covariant bound states\label{WilliamsR_sec:bse}}

\subsection{Bethe--Salpeter equation}
The Bethe--Salpeter equation describing a relativistic bound-state of mass $M$ is
calculated through
\begin{equation}
  \left[\Gamma(p;P)\right]_{tu} = \lambda\! \int\!\!
  \frac{d^4 k}{\left( 2\pi \right)^4}
K_{tu}^{rs}(p,k;P)\left[S(k_+)\Gamma(k;P)S(k_-)\right]_{sr}\,.
\label{WilliamsR_eqn:bse2}
\end{equation}
Here $\Gamma(p;P) \equiv \Gamma^{(\mu)}(p;P)$ is the Bethe--Salpeter
vertex function of a quark-antiquark bound state, specified below. It is
a homogeneous equation, with a discrete spectrum of solutions at momenta
$P^2=-M_i^2$ corresponding to $\lambda\left(P^2\right)=1$. The lightest
of these $M_i$ pertains to the ground state solution. The momenta $k_+ =
k +\eta P$ and $k_- = k-(1-\eta)P$ are such that the total momentum $P$
of the meson is given by $P=k_+-k_-$ and the relative momentum $k=\left(
k_+ +k_- \right)/2$. The momentum partitioning parameter is $\eta$,
of which physical observables are independent. For convenience we choose
it to be $1/2$ without loss of generalisation. The object
$K_{tu}^{rs}(p,k;P)$ is the Bethe--Salpeter
kernel, whose Latin indices refer to colour, flavour and Dirac
structure. 

The Bethe--Salpeter vertex function $\Gamma^{(\mu)}(p;P)$ can be 
decomposed into eight Lorentz and Dirac structures.  The structure is 
constrained by the transformation properties under CPT of the meson
we wish to describe~\cite{LlewellynSmith:1969az}. In particular, our pseudoscalar, scalar and vector
have quantum numbers $J^{P}$ of $0^-$, $0^+$ and $1^-$, respectively. The
axial-vector can be parametrised in two different ways, dependent on
its transformation under charge conjugation,
$J^{PC}=1^{++}$ and $1^{+-}$. 
Taking this into consideration, we can write
down a general basis for each desired meson amplitude,
\begin{equation}
\Gamma_M^{(\mu)}(p;P)=\Bigg\{\begin{array}{cc}
\sum_{i} \phantom{\gamma_5}F_i\left(p;P\right)T^{(\mu)}_i\left(p;P\right) &\,\,\, J^{P} = 0^+,
1^-\\[2mm]
\sum_{i} \gamma_5F_i\left(p;P\right)T^{(\mu)}_i\left(p;P\right) &\,\,\, J^{P}=0^-, 1^+
\end{array}
\end{equation}
where the components $T_i^{(\mu)}$ are given in Table~\ref{WilliamsR_basis}.  For
the axial-vector we point out that for the $J^{PC}=1^{+-}$ ($1^{++}$) state the first (last) four
components significantly dominate and the remainder can be neglected.
Also note that we can impose different charge conjugation properties on
these covariants by changing the odd/evenness of the $F_i\left(p,P\right)$
functions \emph{i.e.} by swapping filled for empty circles in
Table~\ref{WilliamsR_basis}. In this way, states with exotic quantum numbers may
be constructed, see~\cite{Krassnigg:2009zh}.

To give a definite example, the relativistic covariants required to
describe a pion are as follows 
\begin{eqnarray}
\Gamma_\pi(p;P)&=& \gamma_{5}\Big[F_1(p;P)
-i\Pslash F_2(p;P)\nonumber\\[-2mm]\label{WilliamsR_pion}\\[-2mm]
&&\hspace{-9mm}-i\pslash \left(p\cdot P\right)F_3(p;P)
-\left[\Pslash,\pslash\right]F_4(p;P)\Big]\, .\nonumber
\end{eqnarray}
Note, as stipulated in the text, the extra factor of $\left(p\cdot
P\right)$ with
$F_3$ that imposes the even Chebyshev expansion for equal mass
constituents.

\begin{table}[t]
\begin{eqnarray}
\begin{array}{@{}lcc|ccc}
	& & \textrm{Component}      & 0^{-+} & 0^{++}& \\
\hline
\hline
T_1(p;P) &:&  \ONE                  & \circ  & \circ &\\[1mm]
T_2(p;P) &:&  -i\Pslash 		& \circ  & \bullet&\\[1mm]
T_3(p;P) &:&  -i\pslash			& \bullet& \circ&\\[1mm]
T_4(p;P) &:& \left[ \pslash,\Pslash
\right]					& \circ  & \circ&\\[6mm]
& & \textrm{Component}      & 1^{--}  & 1^{++} & 1^{+-} \\
\hline
\hline
T_1^\mu(p;P) &:& i\gamma_T^\mu
&\circ   &\circ  &(\bullet)\\[1mm]
T_2^\mu(p;P) &:& \gamma^\mu_T\Pslash
&\circ   &\bullet&(\circ)  \\[1mm]
T_3^\mu(p;P) &:& -\gamma^\mu_T\pslash +p^\mu_T\ONE
&\bullet &\circ  &(\bullet)\\[1mm]
T_4^\mu(p;P) &:& i\gamma^\mu_T\left[ \Pslash,\pslash
\right]+2ip^\mu_T\Pslash &\circ   &\circ  &(\bullet)\\[1mm]
T_5^\mu(p;P) &:& p^\mu_T\ONE
&\circ   &(\bullet)&\circ  \\[1mm]
T_6^\mu(p;P) &:& i p^\mu_T\Pslash
&\bullet &(\bullet)&\circ  \\[1mm]
T_7^\mu(p;P) &:& -i p^\mu_T\pslash
&\circ   &(\circ)  &\bullet\\[1mm]
T_8^\mu(p;P) &:& p^\mu_T\left[ \Pslash,\pslash \right]
&\circ   &(\bullet)&\circ  \\
\end{array}\nonumber
\end{eqnarray}
\caption{The Dirac structures required to describe mesons of spin $J=0,1$.
For the $J^{PC}=0^{-\pm},1^{+\pm}$ states there is an associated factor of $\gamma_5$ to
  account for parity. Filled circles indicate that 
  the component is multiplied by $\left(p\cdot P\right)$ to impose its
  charge conjugation properties for equal-mass constituents, whereas
  white filled circles indicate there are no additional factors. The axial-vector components 
  in parenthesis are usually neglected. The subscript $T$ 
  indicates transversality with respect to the total momentum.}\label{WilliamsR_basis}
\end{table}

%\subsection{Solving the BS equation}
We solve the BSE for the meson amplitude $\Gamma^{(\mu)}(p;P)$
via matrix methods using the following procedure. First, we project
our BSE onto the scalar amplitudes of our meson decomposition.
This gives rise to either four or eight coupled integral equations for
the $F_i\left(p;P\right)$. To make manifest the angular dependence of the
amplitude functions, we treat the total momentum $P^2$ as a parameter
and expand the function as a series of Chebyshev polynomials in the
angle
$\widehat{p\cdot P} = p\cdot P/|pP|$
\begin{equation}
F_i\left(p;P\right) = \sum_k (i)^k
F^k_i\left(p^2;P^2\right)T_k(\widehat{p\cdot P})\;.
\end{equation}
The functions $F_i^k(p^2;P^2)$ are projected out through use of the
orthonormal properties of the Chebyshev polynomials. With the angular
dependence made explicit, we can evaluate numerically the two
non-trivial angles appearing in the integration measure. We cast the
remaining radial integral in the form of a matrix equation by matching
the external momenta to the radial loop momenta, $p^2_j = k^2_j$ at the
abscissae of our integration nodes. Thus
the amplitude of our BS equation is projected onto the decomposition
$F^k_i\left(p^2_j;P^2\right)$. Schematically we are solving
\begin{equation}
\Gamma = \lambda\, {\bf K}\cdot \Gamma\;,
\end{equation}
for the column vector $\Gamma$ as a parametric equation in $P^2=-M^2$
with $M$ the mass of the meson. A bound-state corresponds to solutions with $\lambda=1$, which
as an eigenvalue problem is equivalent to satisfying 
$\det \left( \ONE-{\bf K} \right)=0$. 
However, since this equation is homogeneous the overall normalisation of the
amplitude is not determined. This is obtained by imposing an auxiliary
condition on the solution, discussed in the next section.

\subsection{Decay constants and normalisation\label{WilliamsR_sec:normalisation}}

\begin{figure*}
  \begin{center}
\begin{eqnarray*}
    \delta^{ij}
	\!\!=\!\!\!&\frac{\partial}{\partial P^2}&\left[ 
	\begin{array}{c}
	\includegraphics[scale=0.6]{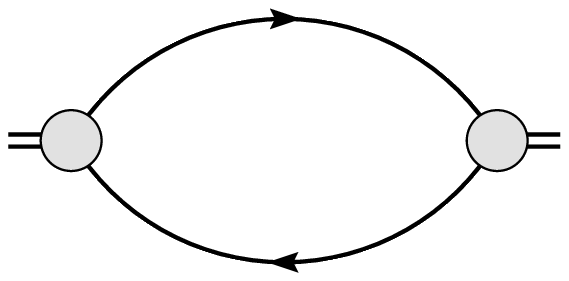}\\
	\end{array}
	+
	\begin{array}{c}
	\includegraphics[scale=0.6]{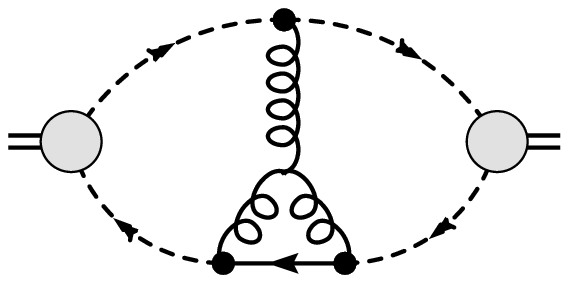}\\
	\end{array}
	%\nonumber\\&&
	+
	\begin{array}{c}
	\includegraphics[scale=0.6]{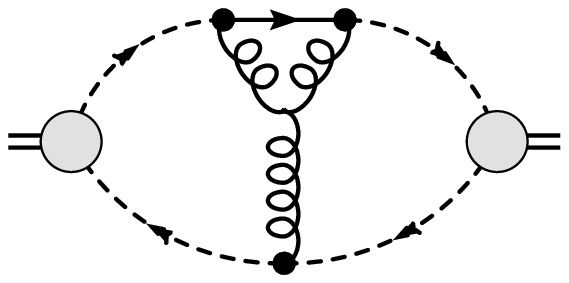}\\
	\end{array}\right.
	\nonumber\\&&
	+\left.
	% %% Abelian diagram
	\begin{array}{c}
	\includegraphics[scale=0.6]{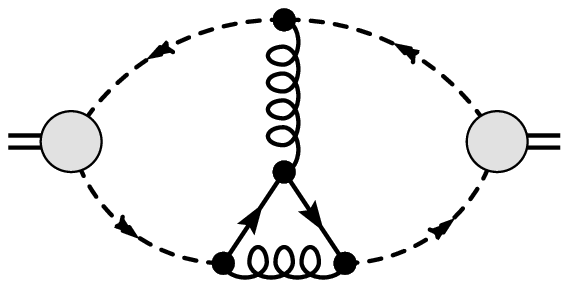}\\
	\end{array}
	%\nonumber\\&&
	+
	\begin{array}{c}
	\includegraphics[scale=0.6]{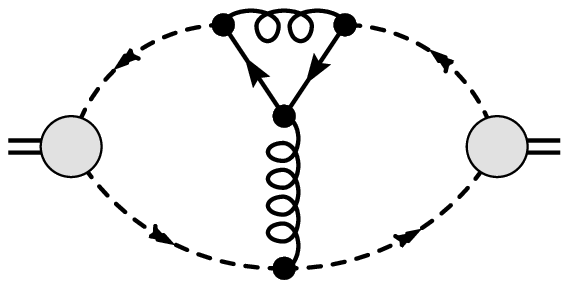}\\
	\end{array}
	+
	\begin{array}{c}
	\includegraphics[scale=0.6]{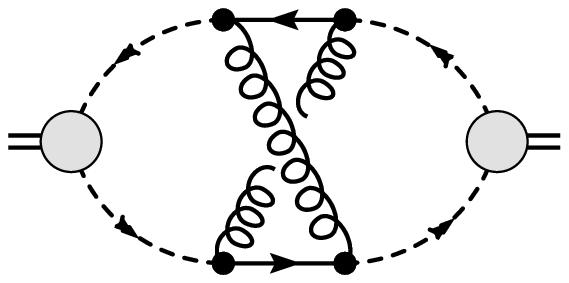}\\
	\end{array}
	\right]
	\nonumber\\
	\end{eqnarray*}
\caption{Normalisation of the Bethe--Salpeter amplitude according to the
Leon-Cutkosky procedure. Dashed lines
represent quark propagators that are kept fixed under action of the
derivative. Numerical factors and the pion exchange diagrams are
suppressed.\label{WilliamsR_fig:cutkoskynorm}}
\end{center}
\end{figure*}

The calculation of observables from the Bethe--Salpter amplitudes, such
as leptonic decay constants for the pseudoscalar and vector
mesons~\cite{Maris:1997hd,Maris:1997tm,Maris:1999nt,Maris:2000sk,Bhagwat:2006pu},
requires that they are properly normalised. This normalisation condition 
is derived by demanding  that the residue of the pole in the four-point 
quark-antiquark Green's function (from which the BSE is derived) is 
unity~\cite{Cutkosky:1964zz,Tandy:1997qf}.  

Using conventions such that $f_\pi=93$ MeV, the standard Leon--Cutkosky
condition reads
\begin{eqnarray}\label{WilliamsR_eqn:cutkosky}
\delta^{ij}&=&2\frac{\partial}{\partial P^2} \tr\int
\frac{d^4k}{(2\pi)^4}\\
& \Bigg[&3 \, \, \bigg(
\overline{\Gamma}_\pi^i(k,-Q)  S(k+P/2)\Gamma_\pi^j(k,Q)S(k-P/2)
\bigg) \nonumber\\
&+&\int \frac{d^4q}{(2\pi)^4} 
[\overline{\chi}_\pi^i]_{sr}(q,-Q)
K_{tu;rs}(q,k;P)[{\chi}_\pi^j]_{ut}(k,Q)\Bigg]\;,\nonumber
\end{eqnarray}
where $Q^2=-M^2$ is fixed to the on-shell meson mass,
the trace is over Dirac matrices and the 
Bethe--Salpeter wave-function $\chi$ is defined by
\begin{equation}
\chi_\pi^j(k;P) = S(k+P/2)\Gamma^j_\pi(k,P)S(k-P/2)\,.
\end{equation}
The conjugate vertex function $\bar{\Gamma}$ is given by
\begin{equation}
\bar{\Gamma}(p,-P)=C \Gamma^T(-p,-P)C^{-1}\;,
\end{equation}
with the charge conjugation matrix $C=-\gamma_2\gamma_4$. For the
rainbow-ladder truncation scheme, which we introduce below, the kernel is
independent of the total momentum $P$ and so vanishes under the
derivative. For more sophisticated truncations this will not be
the case; in Fig.~\ref{WilliamsR_fig:cutkoskynorm} we show the diagrams
that would arise in the truncation scheme that arises from our
discussion of the quark-gluon vertex. These involve the evaluation of three-loop
integrals over non-perturbative quantities. This can be satisfactorily tackled
with the aid of standard Monte-Carlo techniques, with one caveat; the
integration kernel for the crossed-ladder diagram is prohibitively large
after naively taking the Dirac trace. Either we can spend some time
massaging the expression to make it manageable -- which still leaves a
horrific calculation to be done -- or we seek an alternative
normalisation procedure. In fact, such an equivalent normalisation condition
exists in the literature, courtesy of
Nakanishi~\cite{Nakanishi:1965zza}, and was  recently explored
in~\cite{Fischer:2009jm}. Using the momentum dependent eigenvalue
$\lambda(P^2)$ found from solving Eq.~(\ref{WilliamsR_eqn:bse2}) we have the formula
\begin{equation}\label{WilliamsR_eqn:nakanishinorm}
\left( \frac{d\ln(\lambda)}{d P^2}
  \right)^{-1} \!\!\!=\tr\int\nolimits_k \,3\, 
\overline{\Gamma}(k,-P)  S(k_+)\Gamma(k,P)S(k_-)\,.
\end{equation}
For comparative purposes, we tested this truncation against the non-trivial 
kernel's present in Refs.~\cite{Watson:2004kd,Fischer:2008wy}. This is a one-loop expression and consequently requires significantly
less numerical effort to evaluate. We find it simple to apply to all
truncations solved via the homogeneous Bethe--Salpeter equation, such as
the one presented here.

The leptonic decay constant characterising the pion coupling 
to the point axial field is subsequently given by \cite{Tandy:1997qf}
\begin{equation}
f_{\pi}=Z_2\frac{3}{M^2} \tr \int\frac{d^4k}{(2\pi)^4}
\Gamma_\pi(k,-P) \,S(k_+) \,\gamma_5 \,\Pslash \,S(k_-)\;,
\label{WilliamsR_eq:fpi}
\end{equation}
where again the trace is over Dirac matrices, and $k_+=k+P/2$,
$k_-=k-P/2$.   There exist analogous
expressions for the vector mesons~\cite{Maris:1999nt}, where an additional 
factor of $1/3$ must be included due to summation over the polarisation tensor.

One may write a similar equation to (\ref{WilliamsR_eq:fpi}), which corresponds to
the residue of the pseudoscalar vertex:
\begin{equation}
r_{\pi}\!=Z_2 Z_m\, 3\, \tr \!\!\int\!\frac{d^4k}{(2\pi)^4}
\Gamma_\pi(k,-P) \,S(k_+) \,\gamma_5 \,S(k_-)\;.
\label{WilliamsR_eq:rpi}
\end{equation}
The axWTI imposes a relationship between these two residues, known as
the generalised Gell-Mann--Oakes--Renner relation~\cite{Maris:1997hd,Maris:1997tm,Holl:2004fr},
which must hold at and beyond the chiral limit:
\begin{equation}
f_\pi m_\pi^2 = r_\pi\left( m_u(\mu^2) + m_d(\mu^2)
\right)\;,\label{WilliamsR_eqn:gmor}
\end{equation}
where $m_u$, $m_d$ are the masses of the up and down quarks at the renormalisation
point $\mu^2$, and in this work considered to be degenerate. Confirming
this relation serves as a check of both our numerics and indicates how
well our truncation satisfies the axWTI. For the Yang-Mills part of our
truncation this is exactly satisfied, whilst for our pion back-reaction
contribution it is satisfied to better than $1\%$ for pion masses of the order
of $600$~MeV.

\subsection{Constructing a symmetry preserving truncation\label{WilliamsR_sec:cutting}}

In constructing a truncation scheme for our quark DSE and meson BSE, we
must be careful not to break the symmetries of the theory. One of the
most important of these is chiral symmetry. Satisfaction of this ensures
the Goldstone nature of the flavour non-singlet psuedoscalars
irrespective of the details of the model. This is an essential feature
that any dynamical description of light mesons must reproduce, where
explicit breaking due to the introduction of quark masses gives rise to
the physical spectrum of mesons. Explicit and implicit mass dependences
of our interaction should allow for an interpolation between light and 
heavy systems of quarks.

\begin{figure}[h!]
	\begin{eqnarray}
	  \begin{array}{c}
		\includegraphics[scale = 0.40]{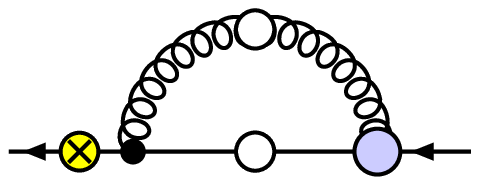}
	  \end{array}
	  \!+\!
	  \begin{array}{c}
		\includegraphics[scale = 0.40]{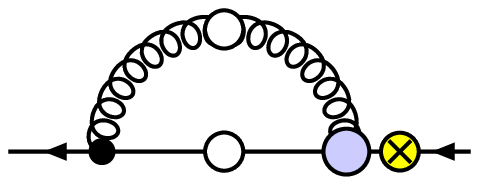}
	  \end{array}
	  =
	  -\!\!\!
	  \begin{array}{c}
		\includegraphics[scale = 0.40]{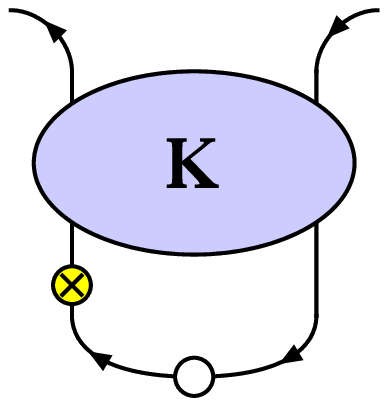}
	  \end{array}
	  \!-\!\!
	  \begin{array}{c}
		\includegraphics[scale = 0.40]{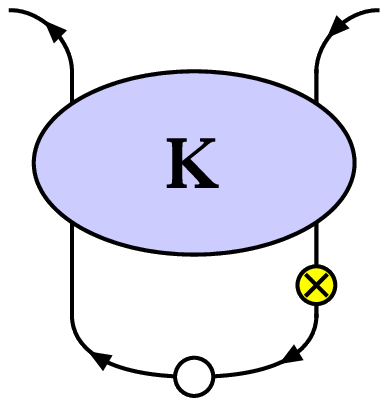}
	  \end{array}\nonumber
	\end{eqnarray}
\caption{The flavour non-singlet axWTI written such that the connection
between the quark self-energy and Bethe--Salpeter kernel is explicit. All
propagators are dressed, with wiggly and straight lines showing gluons
and quarks respectively. The crossed circle indicates the insertion of a
$\gamma_5$.\label{WilliamsR_fig:axwti}}
\end{figure}

The consequences of chiral symmetry can be expressed through the
axial-vector Ward-Takahashi identity for flavour non-singlet mesons
\begin{equation}
  -i P_\mu \Gamma^5_\mu = S_F^{-1}(p_+)\gamma_5 + \gamma_5
  S_F^{-1}(p_-)-2m_R \Gamma^5(p;P)\;,
  \label{WilliamsR_eqn:avwti}
\end{equation}
with $p$ and $P$ the relative and total momentum of the meson
respectively, $p_\pm$ is the combination $p\pm P/2$, $\Gamma^5_\mu$ the axial-vector vertex and $\Gamma^5$ the
pseudoscalar vertex. This equation can be recast in a form that makes
explicit the connection between the quark-gluon vertex and the
Bethe--Salpeter kernel, Fig.~\ref{WilliamsR_fig:axwti}. However, it is not an easy
task to construct a Bethe--Salpeter kernel given a particular truncation
of the quark DSE or arbitrarily specified quark-gluon vertex. Regardless of this
difficulty, progress has been made and constructive schemes do exist in
the literature as we will discuss below.

\begin{figure*}[t!]
\begin{eqnarray*}
	\begin{array}{c}
	\includegraphics[scale=0.7]{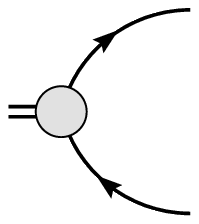}\\
	\end{array}
	&=& 
	\begin{array}{c}
	\includegraphics[scale=0.7]{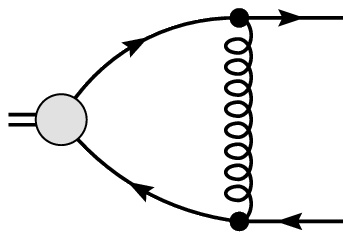}\\
	\end{array}
	+
	\begin{array}{c}
	\includegraphics[scale=0.7]{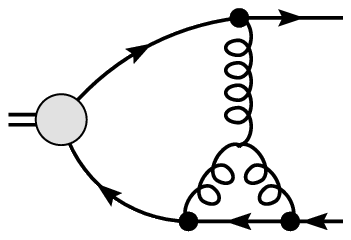}\\
	\end{array}
	+
	\begin{array}{c}
	\includegraphics[scale=0.7]{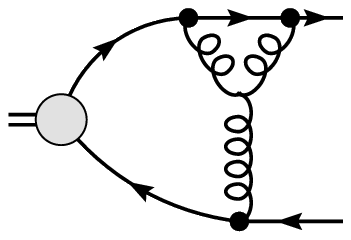}\\
	\end{array}
	+
	\begin{array}{c}
	\includegraphics[scale=0.7]{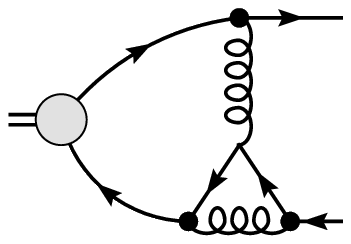}\\
	\end{array}\nonumber\\
	&+&
	\begin{array}{c}
	\includegraphics[scale=0.7]{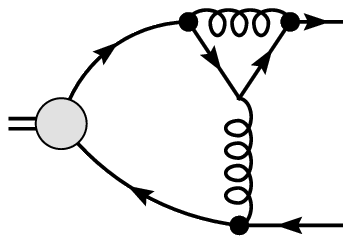}\\
	\end{array}
	+
	\begin{array}{c}
	\includegraphics[scale=0.7]{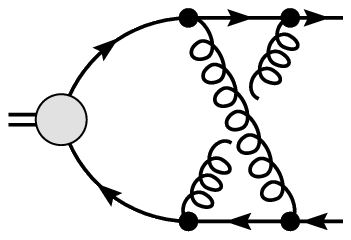}\\
	\end{array}
	+
	\begin{array}{c}
	\includegraphics[scale=0.7]{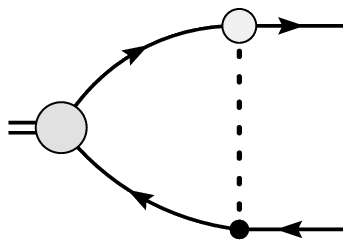}\\
	\end{array}
	+
	\begin{array}{c}
	\includegraphics[scale=0.7]{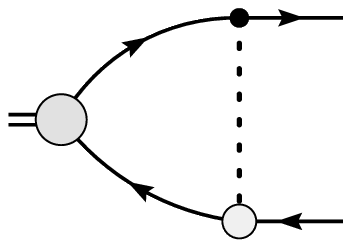}\\
	\end{array}
	\nonumber
	\end{eqnarray*}
\caption{The axWTI preserving BSE corresponding to our vertex truncation. All 
propagators are dressed, with wiggly and straight lines showing gluons
and quarks respectively.\label{WilliamsR_fig:ourbse}}
\end{figure*}

Suppose we choose the simplest possible truncation of the quark-gluon
vertex and replace the (reduced) quark-gluon vertex by a minimally
dressed tree-level vertex 
\begin{equation}
  \Gamma^\mu(q,p) = \Gamma^{\mathrm{YM}}\left( k^2 \right)
  \gamma^\mu\;.
  \label{WilliamsR_eqn:rlvertex}
\end{equation}
Here, $\Gamma^{\mathrm{YM}}$ represents a 
dressing function, dependent only upon the square momentum of the gluon. 
With regards to the quark DSE, this is called the rainbow approximation
due to the rainbow-like appearance of diagrams that it effectively resums. 
The dressing function is chosen to provide a non-perturbative
enhancement of
the quark-gluon vertex and compensates for lost strength from the missing 
basis structures. From the axWTI portrayed in Fig.~\ref{WilliamsR_fig:axwti} 
one finds, by inspection, that the corresponding symmetry preserving 
interaction kernel is
\begin{eqnarray}  
 K_{tu;sr}^{\mathrm{YM}}(q,p;P)
  \!&=&\!
  g^2 Z_{1F}\Gamma^{\mathrm{YM}}(k^2) 
  D_{\mu\nu}\!\left(k^2\right)\!\!
  \left[\frac{\lambda^{a}}{2}\gamma^{\mu}\right]_{ts}\!
  \left[\frac{\lambda^{a}}{2}\gamma^{\nu}\right]_{ru}. \nonumber\\\label{WilliamsR_YMkernel}
\end{eqnarray}
This represents a single gluon exchange, forming the so-called ladder
part of the rainbow-ladder truncation.

Before we consider the constructive scheme for the BS kernel, we make
mention of a different approach followed in Ref.~\cite{Chang:2009zb}.
There, rather than constructing a BS kernel for a given truncation of
the
quark-gluon vertex, they instead constrain the BSE for the meson amplitude itself giving rise to a chiral symmetry
preserving truncation scheme. One caveat, however, is that the quark-gluon vertex
is forced to satisfy the Abelian Ward-Takahashi identity which fixes the
relative strengths of the longitudinal vector and scalar components.
The relative signs and strengths of these terms are important in generating the right spin-orbit
splitting in different meson channels; any deviations from QCD must be
compensated by introducing corrections in different vertex components.
Regardless of these limitations, the relative impact of the individual
transverse vertex components on meson masses can be successfully tested
in this approach and provides for important physical insight.

Since our truncation has a diagrammatic representation, we wish to
employ the constructive scheme to obtain our chiral symmetry preserving
kernel.  It is well-known that one may relate the Bethe--Salpeter kernel $K$ to the
quark self-energy $\Sigma$ by means of the functional 
derivative~\cite{Munczek:1994zz}
\begin{equation}
  K\!\left( x',y';x,y \right) = -\frac{\delta}{\delta
  S\!\left(x,y\right)}\Sigma\!\left(
  x',y' \right)\;,\label{WilliamsR_eqn:cutting}
\end{equation}
a process which is understood to take place in the presence of bilocal
external sources. In momentum space, this procedure can be thought of as
the cutting of internal quark propagators and their replacement by the
Bethe--Salpeter wavefunction, together with the appropriate injection of
the mesons total momentum, $P$. 

Since we include the pion back-reaction according to
Refs.~\cite{Fischer:2007ze,Fischer:2008sp,Fischer:2008wy},
we separate the Bethe--Salpeter kernel into non-resonant Yang-Mills
corrections (YM) and resonant contributions (pion)
\begin{equation}
K_{tu;rs}(p,k;P)= K_{tu;rs}^{\mathrm{YM}}(p,k;P) +
K_{tu;rs}^{\mathrm{pion}}(p,k;P)\;.
\end{equation}
The kernel $K_{tu;rs}^{\mathrm{pion}}(q,p;P)$, consistent with
Eq.~(\ref{WilliamsR_eqn:unquenchdse}) and satisfying the axWTI in the chiral limit is
\begin{eqnarray}
  K_{tu;rs}^{\mathrm{pion}}(q,p;P)
  =
  \frac{1}{4} &\big\{&
[\Gamma^j_{\pi}\left( l_-;p-q\right)]_{ru}[Z_2 \tau^j \gamma_5]_{ts}  \nonumber\\
  &+&
[\Gamma^j_{\pi}\left( l_-;q-p\right)]_{ru}[Z_2 \tau^j \gamma_5]_{ts}  \nonumber\\
  &+&
[\Gamma^j_{\pi}\left( l_+;p-q\right)]_{ts}[Z_2 \tau^j \gamma_5]_{ru}  \nonumber\\
  &+&
[\Gamma^j_{\pi}\left( l_+;q-p\right)]_{ts}[Z_2 \tau^j \gamma_5]_{ru}  \nonumber\\
      &\big\}& D_{\pi}(p-q)\;,
	\hspace{2.5cm}\label{WilliamsR_eqn:kpion}
\end{eqnarray}
where $l_\pm = (p+q\pm P)/2$ and $D_\pi(p) =
1/\left(p^2+M_\pi^2\right)$. The remaining kernel
$K_{tu;rs}^{\mathrm{YM}}(q,p;P)$ is obtained from
Eqs.~(\ref{WilliamsR_eqn:quarkdse},~\ref{WilliamsR_eqn:qg}) via the cutting procedure that
follows from Eq.~(\ref{WilliamsR_eqn:cutting}). This yields the Bethe--Salpeter equation 
portrayed in Fig.~\ref{WilliamsR_fig:ourbse}, where the pion exchange
contributions are indicative of the kernel given above. Note, that to preserve the
appropriate charge conjugation properties we have symmetrised the
kernel.

Now we discuss the different diagrams appearing on the right-hand side of
Fig.~\ref{WilliamsR_fig:ourbse}. The first term is that which arises from the
rainbow-ladder truncation, and is here associated with the inhomogeneous 
term present in the DSE of the quark-gluon vertex.
The next two loop diagrams stem from the dominant non-Abelian corrections 
to the vertex that were considered in Ref.~\cite{Fischer:2009jm}. These appear as
self-energy contributions to the quark-gluon vertex and may be included
without difficulty in the Bethe-Salpter kernel. The only problem is
that these vertices must be calculated for complex incoming and outgoing
quark momenta, but is just an exercise in using appropriate numerical
methods. Introduction of the sub-leading Abelian vertex
correction provides the next three new terms. Two of these are 
again in the form of self-energy contributions to the vertex and can
trivially be added. However, the crossed-ladder kernel deserves further
comment being as it is a significantly more complicated non-planar two-loop
integral. On application of standard projection methods in order to
access the scalar dressing functions of our Bethe--Salpeter amplitude, we
obtain a two-loop integration kernel that most compilers will baulk
at\cite{Vermaseren:2000nd,Reiter:2009ts}. 
There are several different paths one may take to reduce this
complexity. We chose to construct our Bethe--Salpeter kernel in spinor space 
using an explicit matrix representation of the Dirac algebra. This has the 
merit of reducing a million line kernel to a few lines of matrix 
multiplications. It also permits us to, in essence, `uncross' the
gluon-ladder which simplifies the overlapping integration boundary.
The remaining two diagrams of Fig.~\ref{WilliamsR_fig:ourbse} 
correspond to the pion back-reaction kernel, which is given explicitly
in Eq.~(\ref{WilliamsR_eqn:kpion}).

This is the full truncation of the Bethe--Salpeter equation that we
consider here. Only the first diagram comes from the rainbow-ladder
truncation, with everything else classified as `beyond the
rainbow'.  Because we do not trivialize the momentum
dependence of our propagators, the calculation is genuinely two-loop and
we are not restricted as to which meson channels we can investigate.
Moreover, such a truncation allows us to consider the dominant non-Abelian
corrections~\cite{Fischer:2009jm}, together with the sub-leading Abelian 
corrections~\cite{Watson:2004kd}. We can also include hadronic unquenching effects
according to Refs.~\cite{Fischer:2007ze,Fischer:2008sp,Fischer:2008wy}.
We are now in a position to perform a systematic comparison of the
impact of each vertex correction on the spectrum of light mesons.

\section{Results\label{WilliamsR_sec:results}}

\begin{table*}[t!]
	\begin{center}
\renewcommand{\tabcolsep}{1.0pc} % enlarge column spacing
\renewcommand{\arraystretch}{1.1} % enlarge line spacing
\begin{tabular}{@{}c||ccccc}
  Model & $m_\pi$ &  $m_\sigma$ & $m_\rho$  & $m_{a_1}$ & $m_{b_1}$\\
  \hline\hline
  RL      & $138$  &  $645$  & $758$ & $926$ & $912$\\
  NA      & $142$  &  $884$  & $881$ & $1056$ & $973$\\
  AB      & $137$  &  $602$  & $734$ & $889$ & $915$\\
  AB+NA   & $142$  &  $883$  & $878$ & $1052$ & $972$\\
  NA+PI   & $138$  &  $820$  & $805$ & $1040$ & $941$\\
  \hline
  Ref.~\cite{Bhagwat:2004hn} & 132(149)    & 884(997)    & --  & --   & --   \\
  Ref.~\cite{Watson:2004kd}$\dag$ 
  & 138    & 593   & 721   &   --  &  -- \\
  Ref.~\cite{Chang:2009zb}  & 132   & 1060  & --  & --   & --   \\
  \hline\hline
Experiment~\cite{Amsler:2008zzb}   & $138$  & $400$--$1200$ & $776$ & $1230$ & $1230$ \\
\hline
\end{tabular}
\caption{Masses for a variety of mesons calculated
using rainbow-ladder (RL), additional
corrections from the Abelian (AB) and non-Abelian (NA) diagrams, and
with the pion back-reaction (PI). Masses are given in MeV. \newline 
$\dag$ For
Ref.~\cite{Watson:2004kd} we quote our own results as obtained from their 
truncation scheme without use of the real-axis approximation.\label{WilliamsR_tab:results}}
\end{center}
\end{table*}
Using the truncation outlined in the previous sections, we consider the
effect of systematically including different corrections beyond
the rainbow-ladder. Because of simplified form of the gluon
propagator we employ such a comparison is qualitative; the
consistently truncated quantitative study employing Green's functions as
calculated from other DSEs will be reported elsewhere.
However, because the approximations we make do not trivialise the
dynamics and we evaluate the numerics in full, the qualitative
statements we make should be valid in general. An further advantage of using
the gluon propagator of Eq.~(\ref{WilliamsR_eqn:watson}) is that it allows us to 
make a direct comparison with other truncations in the
literature that have employed the same form of the
interaction~\cite{Alkofer:2002bp,Watson:2004kd,Fischer:2009jm,Chang:2009zb}.

\begin{figure}[h]
\includegraphics[width=0.9\columnwidth]{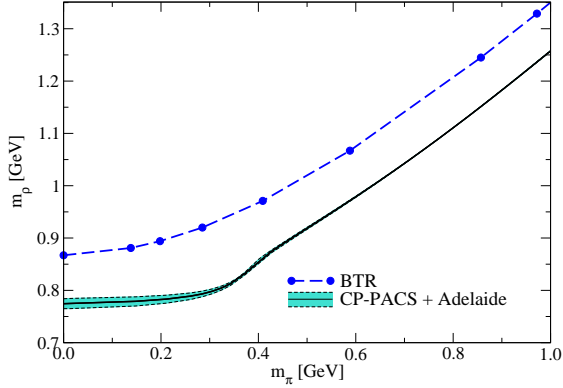}\label{WilliamsR_rhopi}\\
\caption{
$\rho$ mass as a function of the pion mass
(BTR) compared to an extrapolation (CP-PACS + Adelaide)
based on (corrected) lattice data, Ref.~\cite{Allton:2005fb}.
\label{WilliamsR_fig:rhopi}}
\end{figure}

We present results in Table~\ref{WilliamsR_tab:results}. The benchmark for
comparison,
for which the parameters of the interaction are fixed to meson
observables, are those of the rainbow-ladder (RL) approximation. By
including the dominant non-Abelian vertex correction (NA) we find only a
small change in the pion mass, as expected since it is a
pseudo-Goldstone boson. For the remaining bound states we typically see
an enhancement of the mass of the order of $100$--$200$ MeV,
depending upon the meson channel. That is, inclusion of the non-Abelian
diagram is repulsive in the meson channels considered here. In
Fig.~\ref{WilliamsR_fig:rhopi} we show the vector meson mass as a function of the
pseudoscalar mass with the non-Abelian correction included. Note that
the slope will be changed with respect to rainbow-ladder due to the
implicit quark-mass dependence now present in the quark-gluon
interaction.

We compare these results with the study contained in 
Refs.~\cite{Bhagwat:2004hn} (see also \cite{Matevosyan:2006bk}). Therein the authors
employ a resummation of the Abelian vertex and use the
Munczek--Nemirovsky (MN) delta-function to model the gluon. They recognise
that this correction is sub-leading with respect to the non-Abelian diagram
(considered here) and compensate by altering the colour factor accordingly.
However, this gives rise to substantial attraction in the accessible
pseudoscalar and vector mesons, as seen in Table~\ref{WilliamsR_tab:results}.
Since the interaction and parameters are not similar to ours, we quote
their rainbow-ladder results in parenthesis. The precise magnitude of the 
effect is not important, but we see attraction of the order of $5\%$ in
the pseudoscalar and of the order of $100$~MeV in the vector. This is in
contradistinction to the repulsive corrections that we observe and is
seen to be an artefact of the MN interaction or a result of the different
kinematical dependences of the vertices.

If we turn off the dominant
non-Abelian correction and instead consider the sub-leading Abelian
diagram (AB) we obtain the results labelled AB in
Table~\ref{WilliamsR_tab:results}. Based on the relative strength of the
correction due to its colour factors we expect the results to be
$N_c^2$ suppressed, and indeed this is the case.  The mass of the pion
is protected by chiral symmetry and so it receives negligible
contributions from such corrections beyond rainbow-ladder. For the
scalar, vector and $a_1$ axial-vector we see mass reductions of the
order of $20-40$~MeV, while for the $b_1$ there is a slight repulsion of
$3$~MeV. This gives a strong indication that such Abelian corrections to
rainbow-ladder are generally small and attractive.

The result of our Abelian correction can be directly compared with the
study of~\cite{Watson:2004kd}.  In Table~\ref{WilliamsR_tab:results} we
present the results of their truncation using our updated numerical
methods; that is, we calculated the quark propagator for complex values
of the momenta without recourse to the real-value approximation. For 
the pseudoscalar, scalar and vector 
mesons we have qualitatively similar behaviour; they also see attraction in these
meson channels. However, in this hybrid truncation with a
Munczek--Nemirovsky gluon in combination with Eq.~(\ref{WilliamsR_eqn:watson}) we
are not able to find bound-states of the axial-vectors. Instead, our
quark propagator develops non-analyticities in the complex plane
which restrict our bound-state calculation to masses below $800$~MeV.
If we consider the original results of Ref.~\cite{Watson:2004kd} for the
axial-vectors, a $300$~MeV repulsion was observed in the $b_1$ channel,
compared with our $3$~MeV repulsion in the full two-loop calculation.
This gives a strong indication that use of the Munczek--Nemirovsky
delta-function gluon can give rise to large model artefacts. This gives
rise to a degree of uncertainty in any qualitative predictions made.

Now, if we consider both the Abelian and non-Abelian corrections to the
quark-gluon vertex together we might expect the effect on the meson
bound-state to stack. That is, for the vector meson we would expect to see a
$\sim120$ repulsion from the NA correction, with a $\sim 25$ attraction
from the AB correction, resulting in a bound-state mass of $\sim 860$. 
Instead, we see that the Abelian
corrections are heavily suppressed by the non-Abelian ones for all meson channels, giving
results that are almost identical to the those from the $NA$ diagram
alone. The conclusion is that naively adding and subtracting the results 
of different independent studies without performing the combined dynamical 
calculation can be misleading. It is thus reasonable, with the
interaction model presented here, to ignore the Abelian diagram completely 
whenever we take the dominant non-Abelian diagram into account.

Finally, having demonstrated that the Abelian diagram is significantly
suppressed in the meson spectrum, we consider the dominant non-Abelian 
corrections beyond rainbow-ladder with
unquenching effects in the form of pion exchange (PI). This gives rise
to the fifth row of Table~\ref{WilliamsR_tab:results}. As expected, the inclusion of a
pion exchange kernel is generally attractive, with an $80$ MeV reduction
of the vector meson mass with respect to the RL+NA result. This
demonstrates, as suspected in~\cite{Bender:1996bb,Fischer:2009jm} the near
cancellation of beyond-the-rainbow corrections with unquenching effects in this
channel. Here, this is not an exact mechanism but the result of
dynamical combinations of attractive and repulsive components of the
quark-gluon vertex; a more accurate picture will be revealed when
improved approximation schemes are employed that permit quantitative
study.

Finally, we make a comparison with the truncation of
Ref.~\cite{Chang:2009zb}. Rather than extend the rainbow-ladder
truncation by diagrammatically representable corrections to the
quark-gluon vertex, they constrain the longitudinal part of the
quark-gluon vertex with the aid of the Abelian Ward-identity. This is
equivalent to writing the quark-gluon vertex as
\begin{equation}
  \Gamma^\mu(k,p) = \Gamma_{\mathrm{NA}}\left(  q^2\right)
  \Gamma^\mu_{\mathrm{BC}}(k,p)\;,
  \label{WilliamsR_eqn:nonbc}
\end{equation}
which is the Ball-Chiu Ansatz as obtained in QED, combined with a
dressing function that provides the non-Abelian character of the vertex.
Note that such a separable ansatz has been extensively used in studies
of the quark-propagator~\cite{Fischer:2003rp,Alkofer:2003jj}. Their results, taken from the figure
contained in Ref.~\cite{Chang:2009zb}, are shown in
Table~\ref{WilliamsR_tab:results}. This ansatz has the virtue of providing a
significant scalar contribution to the quark-gluon interaction that will
prove useful in the study of mesons containing significant orbital
angular momentum.

%%%%%%%%%%%%%%%%%%%%%%%%%%%%%%%%%%%%%%%%%%%%%%%%%%%%%

\section{Discussion and Outlook\label{WilliamsR_sec:outlook}}
We improved upon previous beyond-the-rainbow investigations, which
focused upon pion unquenching
corrections~\cite{Fischer:2007ze,Fischer:2008sp,Fischer:2008wy} and the
dominant non-Abelian corrections to the quark-gluon
vertex~\cite{Fischer:2009jm} by including the sub-leading Abelian
corrections to the quark-gluon vertex. This enables us to make a direct
comparison with existing works in the literature that employ
trivial~\cite{Bender:1996bb,Bender:2002as,Bhagwat:2004hn,Matevosyan:2006bk} and finite
width~\cite{Watson:2004kd,Watson:2004jq} ans\"atze for the gluon propagator. We
investigated the impact of these corrections beyond-rainbow ladder
independently and in combination, finding that the Abelian-like
corrections are further suppressed relative to the non-Abelian ones in
dynamical calculations. This suggests that naively adding and
subtracting the relative mass shifts from separate investigates prove to
be misleading. We also found that the MN delta-function ansatz
gives can give rise to spurious results and model artefacts, thus
rendering many conclusions with these models qualitatively unreliable. By including 
the pion back-reaction in combination with the leading non-Abelian 
vertex correction we see that there is a degree of cancellation between the two 
corrections in the vector meson channel.

The study we presented involves many technical developments in the
calculation of the quark propagator and quark-gluon vertex at complex
momenta. For highly non-trivial two-loop diagrams such as the
crossed-ladder kernel, new and improved solution methods were required
to render the calculation tractable. Despite these developments, the
investigation remains qualitative and serves only as an exploratory study of the
relevance of the different contributions beyond rainbow-ladder
considered thus far.

Future investigations will focus upon including realistic
input from the ghost and gluon sector of the theory~\cite{Fischer:2008uz}, 
appropriate dressings the internal vertices contained within
the Abelian and non-Abelian corrections~\cite{Alkofer:2008tt} and with input
provided from the DSE of the three-gluon vertex~\cite{Alkofer:2008dt}. An extension
to strange and charm quarks is desirable and possible within our scheme. These 
are currently under investigation and the results will be reported elsewhere. 

In addition to choosing a more theoretically motivated truncation
scheme, it would be interesting to introduce further unquenching
effects to our Bethe--Salpeter kernel and also our gluon propagator. 
Such sophisticated studies such as these are left for the future.

\section{Acknowledgements}\label{WilliamsR_sec:ack}
RW wishes to thank Christian S. Fischer for a careful reading of this
manuscript and contributions to many aspects of this work, and also
Gernot Eichmann, Christian Kellermann and Peter Watson for useful discussions. 
This work was supported by the Helmholtz-University Young Investigator 
Grant No. VH-NG-332.

\end{document}